# DESIGN OF ENGINEERED LIVING MATERIALS FOR MARTIAN CONSTRUCTION

**Nisha Rokaya**

Department of Plant Pathology, University of Nebraska–Lincoln, Lincoln, NE 68583

**Erin C. Carr**

School of Biological Sciences, University of Nebraska–Lincoln, Lincoln, NE 68588

**Richard A. Wilson**

Department of Plant Pathology, University of Nebraska–Lincoln, Lincoln, NE 68583

**Congrui Jin\***

Engineering Technology and Industrial Distribution, Texas A&M University, College Station, TX 77843

\* Corresponding author. E-mail addresses: jincongrui@tamu.edu

**ABSTRACT**

As the next step in extraterrestrial exploration, many engineers and scientists around the country revealed their intense interest to enable multiplanetary human life, including colonizing Mars. This study proposes that architecture on Mars can be realized by a synthetic lichen system, composed of diazotrophic cyanobacteria and filamentous fungi, which produce abundant biominerals and biopolymers to bond Martian regolith into consolidated building blocks. These self-growing building blocks can be assembled into a wide range of structures. Diazotrophic cyanobacteria will 1) fix carbon dioxide and dinitrogen from the atmosphere and convert them into oxygen and organic carbon and nitrogen sources to support filamentous fungi; and 2) give rise to high concentrations of carbonate ions because of photosynthetic activities. Filamentous fungi will 1) bind metal ions onto fungal cell walls and serve as nucleation sites to promote biomineral precipitates; and 2) assist the survival and growth of cyanobacteria by providing them water, minerals, additional carbon dioxide, and protection. This report presents the major progress of the project. It has been tested and confirmed that such co-culture systems can be created, and they grow very well solely on Martian regolith simulants, air, light, and an inorganic liquid medium without any additional carbon or nitrogen sources. The cyanobacterial and fungal growth in such co-culture systems is significantly better than their axenic growth due to mutual interactions. The amounts and morphologies of the precipitated crystals vary remarkably depending on the cultivation condition.

## 1. INTRODUCTION

As the next step in extraterrestrial exploration, many engineers and scientists around the country revealed their intense interest to enable multiplanetary human life, including colonizing Mars. Without the presence of any antecedent, the first Martian architecture is typically portrayed as exquisite industrial prefabricates, as shown in Fig. 1(a). However, the challenges of reaching such a distant realm, transporting entire sets of building parts, and assembling and maintaining them in environments that are far more challenging than even the most extreme environments on Earth are well beyond our current science and technology.

The goal of this project to bring in game changers – radical changes to the state-of-the-art construction technologies such that the construction process can adjust to the restricted resources and adverse environments on new planets. This report summarizes the main progress of a recent study, which aims to engineer microorganisms to create biomaterials that glue Martian regolith particles into structures, as shown in Fig. 1(b).

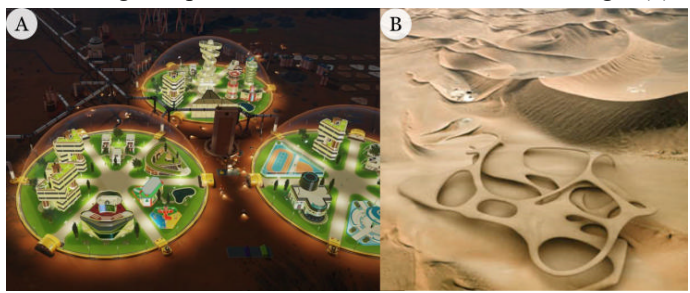

**Figure 1.** (a) A Google image search for "Mars colony" returns a lot of hi-tech structures. (b) This study proposes that microorganisms can be engineered to create abundant biomaterials that will glue Martian regolith particles into bio-structures.

A two-component synthetic lichen system is created, as shown in Fig. 2(a). Diazotrophic cyanobacteria are in charge of 1) fixing carbon dioxide and dinitrogen from the atmosphere and converting them into oxygen and organic carbon and nitrogen sources to support filamentous fungi, as shown in Fig. 2(b); and 2) causing high concentrations of carbonate ions because of photosynthetic activities, which is an essential process for biomineral precipitation. Filamentous fungi are in charge of 1) attracting metal ions onto fungal cell walls and serving as nucleation sites to promote biomineral precipitates; and 2) assisting the survival and growth of cyanobacteria by providing them water, minerals, additional carbon dioxide, and protection. Both components secrete biopolymers that enhance the adhesion and cohesion among Martian regolith and precipitated particles.

## 2. BACKGROUND AND MOTIVATION

To realize in-situ utilization of Martian regolith, various methodologies for bonding regolith particles have been proposed. Scott et al. bonded Martian regolith simulants using a magnesium-based cement [1]. It has been found that magnesium oxide, which needs to be extracted from magnesium carbonates on Mars, can react with amorphous silica to form magnesium-silica-hydrate, which functions like the calcium-silicate-hydrate formed in ordinary Portland cement. Wan et al. mixed molten sulfur with Martian regolith simulants to produce sulfur concrete [2]. Sulfur needs to be extracted from gypsum or pyrite minerals on Mars. Alexiadis et al. created and tested geopolymer concrete using Martian regolith simulants [3]. Geopolymerization hinges on the activation of regolith using a variety of activators and thus its feasibility requires the production of activator materials on Mars. In summary, all these approaches require significant amounts of human assistance, and thus cannot combat the obvious challenge of manpower shortage on Mars.

On the other hand, microbe-mediated self-growing technology has been intensely investigated during the past decades. Dosier et al. utilized bacterial biomineralization to consolidate sand particles into masonry units [4]. She founded a company called bioMASON based on her patented process [5]. Besides biominerals, fungal mycelium has also been utilized as bonding materials. A company called Ecovative Design was founded based on its patented process [6]. Another company, MycoWorks, also uses fungal mycelium to create building blocks and leather-like fabrics. For the application of space exploration, Indian Institute of Science demonstrated the feasibility of using ureolytic bacteria to promote the production of calcium carbonate to consolidate Martian regolith simulants into bricks [7]. In the United States, National Aeronautics and Space Administration (NASA) Ames Research Center has an ongoing project entitled "Myco-Architecture off Planet", which explores the feasibility of using fungal mycelium to bond Martian regolith simulants [8].

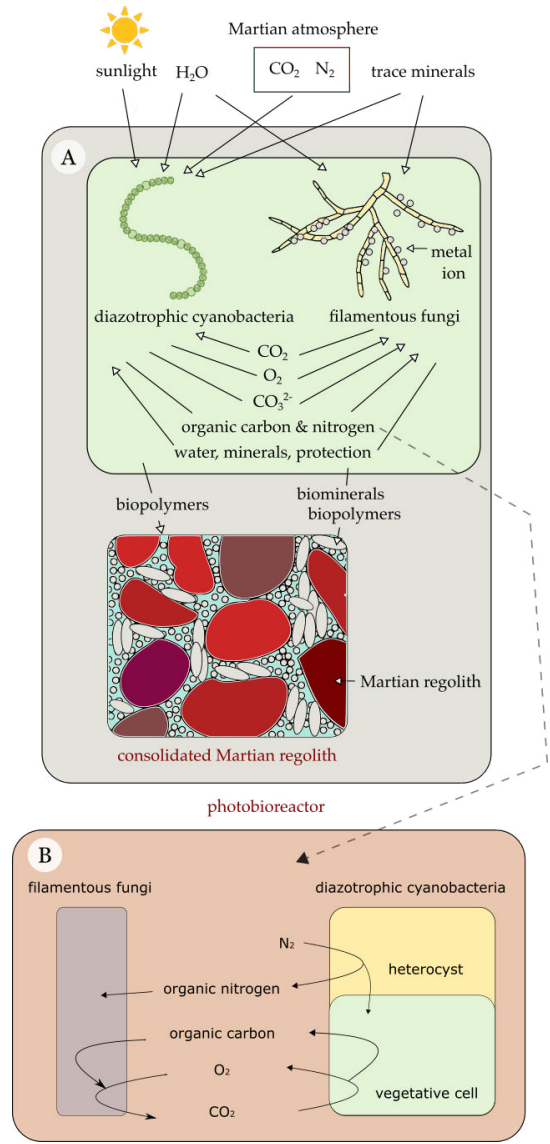

**Figure 2.** (a) Schematic illustration of the self-sustaining system. (b) Diazotrophic cyanobacteria are responsible for fixing carbon dioxide and dinitrogen from the atmosphere and converting them into oxygen and organic carbon and nitrogen sources to support filamentous fungi.

## 3. HYPOTHETICAL BASIS

In this project, heterotrophic filamentous fungi, instead of heterotrophic bacteria, are selected to fit the role of bonding material producers. This idea is based on the following two hypotheses. The first one is that filamentous fungi are better candidates than bacteria as bonding material producers due to their remarkable survivability against harsh conditions and extraordinary ability to promote large amounts of biomineral precipitates. When investigating microbe-based self-healing concrete, Zhang et al. observed that fungi-based healing time could reach 11.3% of the time required by bacteria [9].

Experimental results demonstrated that filamentous fungi could promote much larger amounts of calcium carbonate precipitates within the same time duration compared with bacteria [10].

The second hypothesis is that a synthetic lichen system can be created by pairing heterotrophic filamentous fungi with photoautotrophic diazotrophic cyanobacteria that exhibit mutualistic interactions. Natural lichens represent one of the most widespread phototroph-heterotroph symbioses. Although investigation and utilization of natural lichens are currently minimal, synthetic co-culture systems have been explored by several pioneering studies. Li et al. [11] constructed a self-supporting synthetic symbiosis consisting of the sucrose-secreting cyanobacterium cscB$^+$ *Synechococcus elongatus* PCC 7942 and the yeast *Rhodotorula glutinis*, whose growth purely relies on the presence of light, carbon dioxide, and an inorganic liquid medium. It was observed that the phototroph was able to provide organic compounds to support the heterotroph whereas the heterotroph assisted in phototroph survival and robust growth. Jiang et al. [12] also constructed a stable synthetic partnership consisting of the cyanobacterium *Nostoc* PCC 6720 and the filamentous fungus *Aspergillus niger*, growing on light and carbon dioxide in an inorganic liquid medium. These proof-of-concept studies have demonstrated significant potential of creating mutualistic partnerships between photoautotrophic cyanobacteria and heterotrophic fungi as a stable platform for the self-sustained production of biomaterials. Note that none of the previous studies were conducted in the environment of Martian regolith simulants.

## 4. MATERIALS AND METHODS
### 4.1. Collection of Alkaliphilic Strains

In the past years, new Martian regolith simulants kept being developed, among which MGS-1 was fabricated based on individual components to match the mineralogy of Martian regolith [13], making it so far the most accurate basaltic Martian regolith simulant. The pH of MGS-1, being approximately 9.5 [14], remains the most significant challenge for this project since this high pH inhibits the growth of most microorganisms. Thus, alkaliphilic cyanobacteria and fungi need to be obtained from natural environments or via genetic editing.

Wild-type isolates typically possess exceptional resistance to temperature fluctuations and superior adaptation to nutrient-poor and dry habitats. In the past few years, grass root-associated fungi were collected at the Nottingham Serpentine Barrens in Pennsylvania, the Pine Barrens in New Jersey, and the Rocky Mountains of Alberta in Canada, respectively, where the soils have high pH and low nutrients. Fig. 3 shows the field collection conducted at Nottingham County Park, which is part of the Nottingham Serpentine Barrens.

Some wild-type isolates are found to possess superior ability to promote large amounts of precipitates, but they do not possess sufficient survivability against the alkaline environment of MGS-1. Genetic editing is utilized to address this issue. For filamentous fungi, alkaline pH is registered by a signal transduction pathway involving the products of at least six pal genes and transmitted to a transcription factor encoded by the *pacC* regulatory gene. The *pacC$^c$* mutations in the *pacC* regulatory gene leads to alkalinity mimicry [15].

Non-photosynthetic bacteria survive high pH using a wide range of mechanisms, among which the best studied approach is the increased expression and activity of monovalent cation/proton antiporters [16]. These transmembrane proteins regulate the intracellular pH via the uptake of protons. A variety of such antiporters are recognized in cyanobacteria [17-19]. Their role in pH homeostasis can be investigated by gene knockout studies.

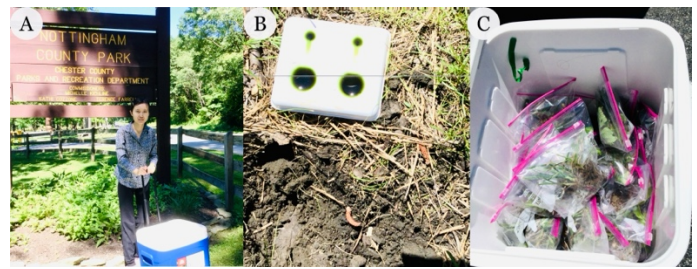

**Figure 3. (a) Field collection at Nottingham County Park. (b) Soil samples were collected and testing for pH. The surface pH was neutral and rose to about 8.0 to 9.5 at the 20-inch depth. (c) For each site, 10 apparently healthy plant individuals were collected at a distance of at least 10 meters apart. Samples were processed within 24 hours of collection and kept on ice during transportation.**

### 4.2. Survivability and Mutualism Testing

The collected fungal strains first go through survivability testing. The strains that demonstrate optimal survivability in the growth medium containing Martian regolith simulants and potato dextrose agar (PDA) are selected for the second testing, i.e., the mutualism testing. A series of co-culture systems are created, and their growth solely based on Martian regolith simulants, air, light, and an inorganic liquid medium, i.e., a modified Bold's Basal Medium (BBM) [20], with or without additional carbon or nitrogen sources is observed. For each monoculture or co-culture, four different cases are tested, including C+N+, C+N-, C-N+, and C-N-.

To assess microbial growth in each well, three quantitative characterization methods are applied, i.e., resazurin assay, fungal plating on selective medium, and measurement of phycocyanin autofluorescence [20]. The co-culture systems with the optimal survivability and mutualism without any additional carbon or nitrogen sources are selected as candidates for Martian construction. The detailed protocol was presented by Rokaya et al. [20].

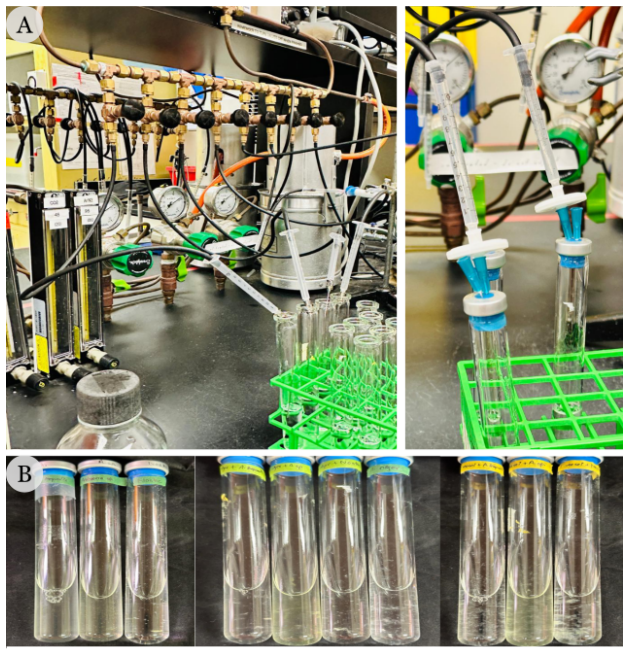

**Figure 4. (a)** Survivability testing is performed using a laboratory-scale photobioreactor. **(b)** Some co-culture systems demonstrate robust growth solely based on the presence of light, water, carbon dioxide, dinitrogen, and trace minerals extracted from Martian regolith simulants.

Neither cyanobacteria nor fungi could directly thrive on Mars. The Martian atmosphere has a pressure of 7.5 mbar, only 0.7% of that of Earth, which does not support the metabolism of most microbes. Since there exists only 2.8% dinitrogen in the Martian atmosphere, cyanobacteria cannot extract enough nitrogen to support their diazotrophic growth. In addition, the surface temperature on Mars shows extreme diurnal fluctuations from 174 K to 295 K, unsuitable for microbial growth. Finally, the absence of an ozone shield means that radiation protection is needed. Hence, a photobioreactor is needed to provide the microbes with protection against damaging radiation and optimal growth condition.

Survivability and mutualism testing are performed within a laboratory-scale photobioreactor, as shown in Fig. 4(a). Except for the small quantities of cyanobacterial and fungal spores, all the inputs for the photobioreactor can be directly found on Mars, i.e., water, carbon dioxide, dinitrogen, and Martian regolith. Some co-culture systems demonstrate robust growth solely based on the presence of light, water, carbon dioxide, dinitrogen, and trace minerals extracted from Martian regolith simulants, as shown in Fig. 4(b).

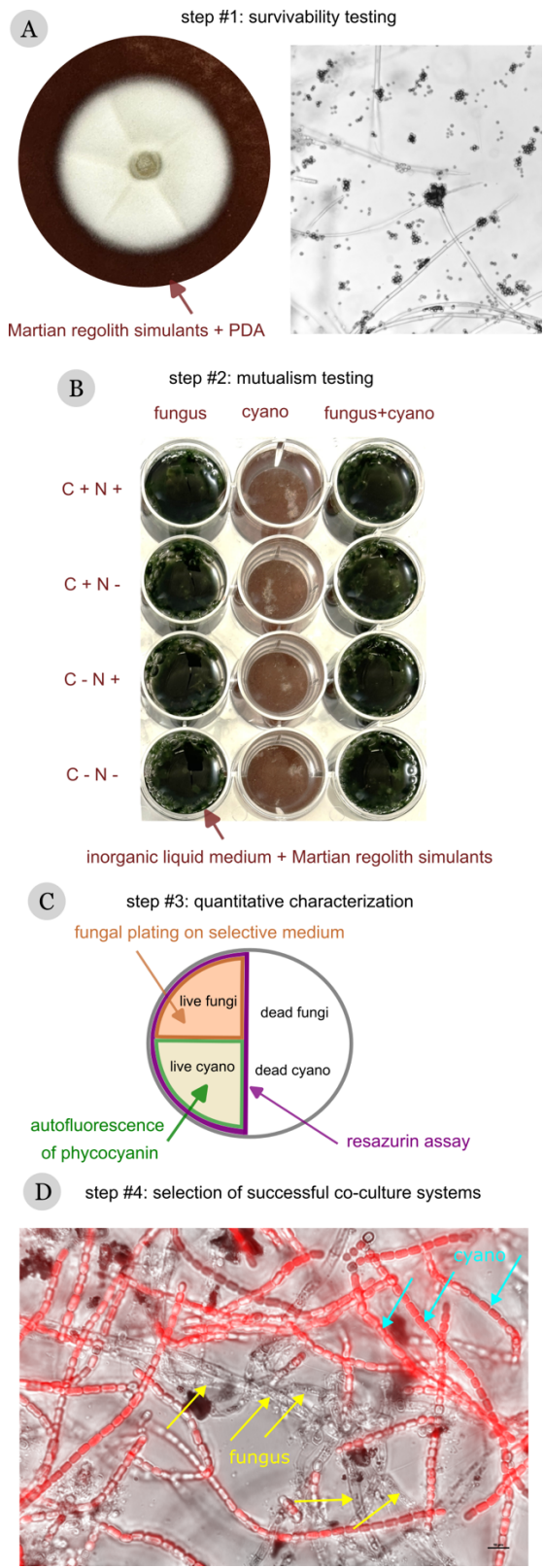

**Figure 5.** Through a four-step investigation, synthetic symbioses consisting of diazotrophic cyanobacteria and filamentous fungi were constructed and their successful growth on Martian regolith simulants, air, light, and an inorganic liquid medium without any additional carbon or nitrogen sources have been tested and confirmed.

### 4.3. Metabolic Interactions During Co-Culturing

To understand if the cyanobacterial component provides nutrients for its fungal partner in the C-N- co-cultures, metabolomics is an essential tool to study the differences in metabolites between monocultures and co-cultures. Therefore, the metabolites from monocultures and co-cultures, both grown solely on Martian regolith simulants, air, light, and an inorganic liquid medium without additional carbon or nitrogen sources, are of utmost importance for investigation.

In this study, Benedict's test is used to identify reducing sugars present in the supernatant of each well. 1 mL of supernatant is withdrawn from each well and mixed with 2 mL of Benedict's reagent. The solution is then heated in a boiling water 5 minutes. Upon mixing and heating of reducing sugars and Benedict's reagent, a reduction reaction occurs, and a change in color can be observed. The degree of the color change from blue to brown reflects the percentage of reducing sugars present in the supernatant of each well. After the samples cool for 5 minutes, optical density is measured at 600 nm using a spectrophotometer.

### 4.4. Material Characterization of Biomaterials

Due to the complexity of the biomineralization process, material characterization technologies are employed to understand the effect of various influencing factors on the production of biomaterials, such as fungal strain/species, cyanobacterial strain/species, atmospheric composition, illumination cycle, temperature range, cultivation duration, and growth medium composition, etc.

To monitor the structural and morphological changes of the solid precipitates over time, scanning electron microscopy (SEM) and energy dispersive X-ray spectroscopy (EDS) are performed at the early, intermediate, and terminal stages of each mutualism test. SEM and EDS provide a three-dimensional overview of the morphological and composition evolution of the solid precipitates.

To test the samples in each well, the growth medium is removed from the culture and the samples are washed by immersing twice in 70% ethanol for 15 minutes. Representative samples are then spread as a thin layer on cover slips and left overnight to dehydrate with 95% ethanol. The samples are then completely dried in an oven at 40°C for 2 days. The dried samples are mounted onto aluminum stubs using double-sided carbon tape and sputter-coated with a 5 to 10 nm layer of chromium to ensure electrical conductivity. The crystals are observed using a Hitachi S4700 Field-Emission SEM and an FEI Helios NanoLab 660 SEM at accelerating voltages between 5 kV and 20 kV. EDS is performed by an EDAX Octane Elect EDS system.

## 5. RESULTS AND DISCUSSION
### 5.1. Survivability and Mutualism Results

Through a four-step investigation as shown in Fig. 5, it can be concluded that the co-culture systems that demonstrate optimal survivability and mutualism without any additional carbon or nitrogen sources include six stable symbioses consisting of a diazotrophic cyanobacterium *Anabaena* sp. (UTEX2576) and a filamentous fungus *Penicillium sipitatus* (ATCC10500), *Aspergillus niger* (ATCC16888), *Trichoderma reesei* (ATCC13631), *Trichoderma viride* (HF3TV), *Trichoderma viride* (HF3MP), or *Aspergillus flavus* (ATCC9643). Their successful growth on Martian regolith simulants, air, light, and an inorganic liquid medium without any additional carbon or nitrogen sources have been tested and confirmed. In addition, based on the testing results within the photobioreactor, the co-culture of *Anabaena* sp. and *A. niger* and the co-culture of *Anabaena* sp. and *T. reesei* demonstrate robust growth solely based on the presence of light, water, carbon dioxide, dinitrogen, and trace minerals extracted from Martian regolith simulants.

The autofluorescence of phycocyanin characterizes the number of live cyanobacterial cells, and the results after the 45-day incubation period are shown in Fig. 6. The resazurin assay assesses the number of live cells, both fungal and cyanobacterial, and the results after the 45-day incubation period are shown in Fig. 7. In general, the cyanobacterial or fungal growth in the co-culture systems is much better than their axenic growth, demonstrating the importance of mutual interactions.

The method of fungal plating on selective medium assesses the number of live fungal cells in each well, and the results after the 45-day incubation period are shown in Fig. 8. 100 μL of homogenous sample in each well is withdrawn and spread evenly on a selective medium in a Petri dish. The colony surface area is estimated after 3 days of incubation, indicating that there are significant amounts of live fungal cells after 45 days of incubation, even when no supplemented carbon or nitrogen source is present.

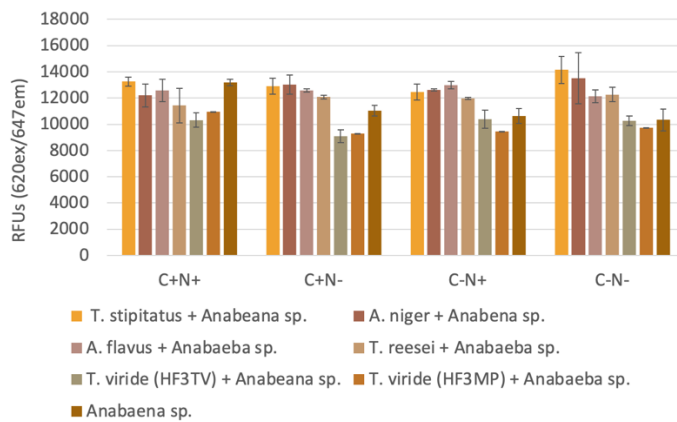

**Figure 6. Results of phycocyanin autofluorescence measurement after 45-day incubation, characterizing the number of live cyanobacterial cells.**

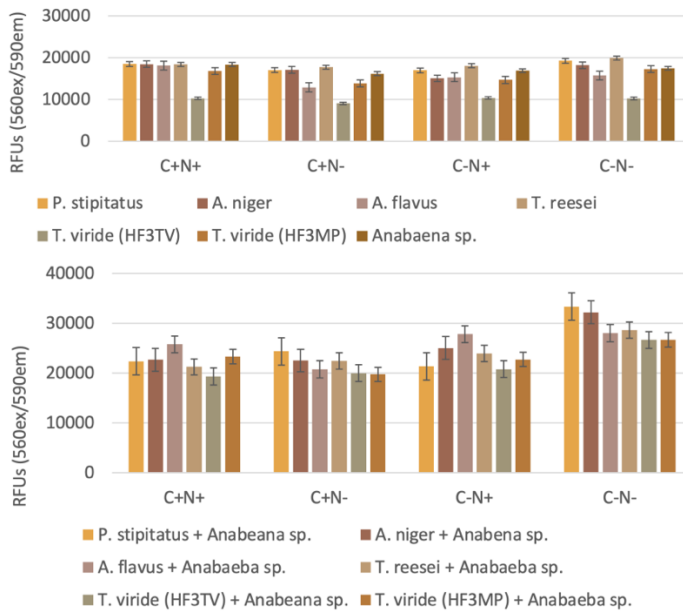

**Figure 7.** Results of resazurin assay after 45-day incubation, characterizing the number of live cells, both fungal and cyanobacterial.

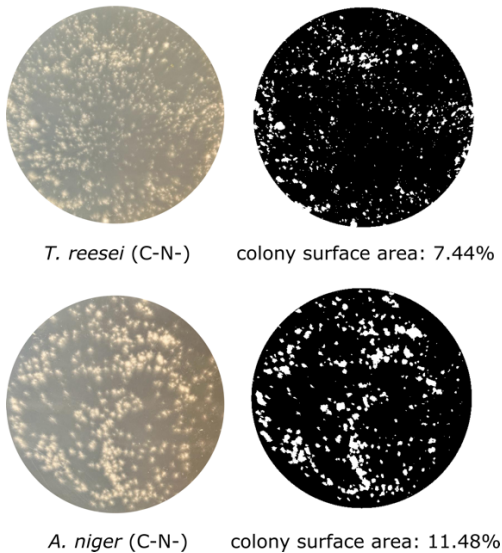

**Figure 8.** Results of fungal plating on selective medium after 45-day incubation, assessing the amount of live fungal cells.

## 5.2. Metabolic Interactions During Co-Culturing

Benedict's test is used to identify reducing sugars in the supernatant of monoculture and co-culture wells, both grown solely on Martian regolith simulants, air, light, and an inorganic liquid medium without additional carbon or nitrogen sources. The degree of the color change reflects the percentage of reducing sugars present in the supernatant of each well, as shown in Fig. 9. The change in color in the co-culture supernatants indicates that the cyanobacteria are producing reducing sugars, not initially provided in the media, for its fungal partner.

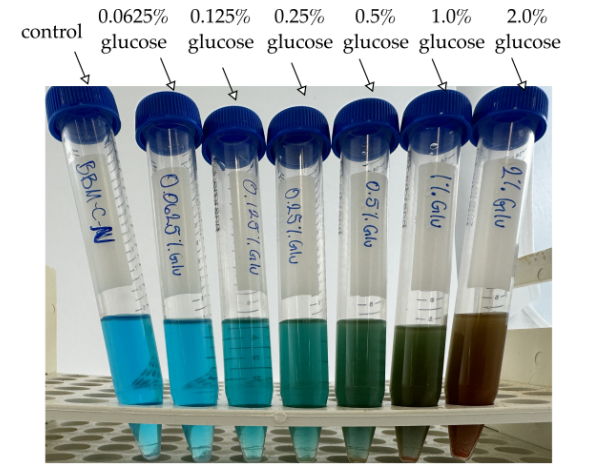

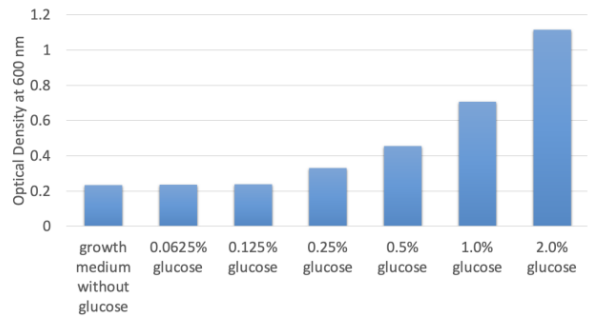

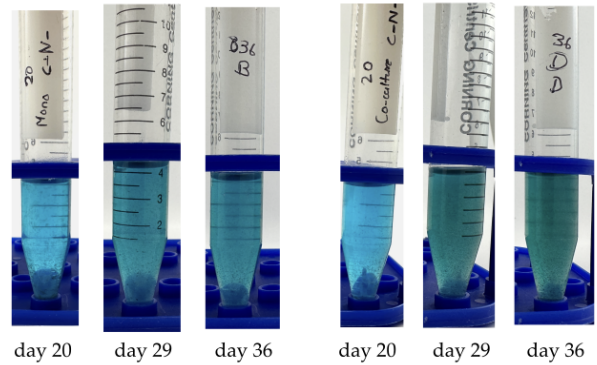

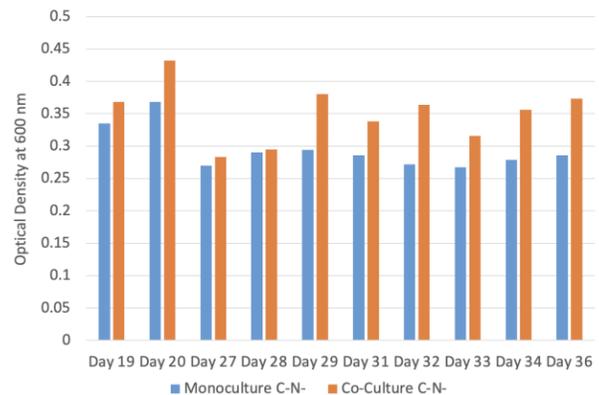

**Figure 9.** Results of Benedict's test used to identify reducing sugars present in the supernatant of cyanobacterial monoculture wells and co-culture wells.

### 5.3. Characterization Results of Biomaterials

It has been observed that the cultivation condition could significantly influence the precipitation process, and even under the same cultivation condition, the amounts and morphologies of the precipitated crystals vary remarkably depending on the fungal component, as shown in Fig. 10.

When the cyanobacterial component is fixed to be *Anabaena* sp., *T. reesei* precipitates prism-like crystals, *T. viride* (HF3MP) precipitates plate-like crystals, *T. viride* (HF3TV) precipitates rod-like crystals, *P. sipitatus* precipitates crystals with gel-like structures, and on the other hand *A. niger* does not precipitate large amounts of crystals but its mycelial network shows notable growth. EDS analysis showed that the crystals are mainly calcium carbonate. In the microbe-free control medium, no crystal growth is observed.

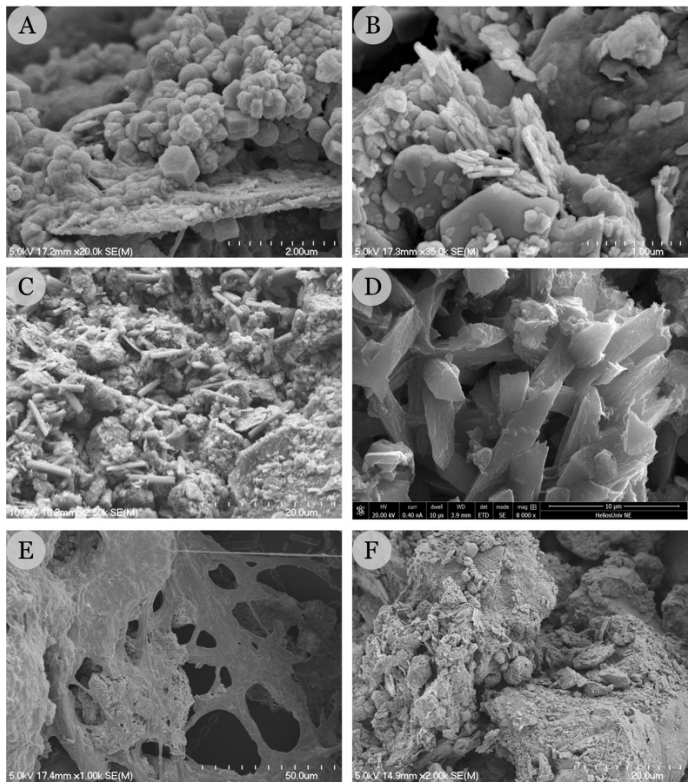

**Figure 10.** Representative SEM images of precipitated crystals taken after 14-day incubation from the C-N- co-culture wells whose fungal component is (a) *T. reesei*, (b) *T. viride* (HF3MP), (c) *T. viride* (HF3TV), (d) *P. sipitatus*, (e) *A. niger,* and (f) the microbe-free control growth medium.

## 6. CONCLUSION AND FUTURE WORK

In this study, synthetic symbioses consisting of diazotrophic cyanobacteria and filamentous fungi were constructed and their successful growth on Martian regolith simulants, air, light, and an inorganic liquid medium without any additional carbon or nitrogen sources have been tested and confirmed. It has been observed that the cultivation condition could significantly influence the precipitation process, and even under the same cultivation condition, the amounts and morphologies of the precipitated crystals vary remarkably depending on the fungal component. The potential of this self-growing technology in enabling long-term extraterrestrial exploration and colonization is significant.

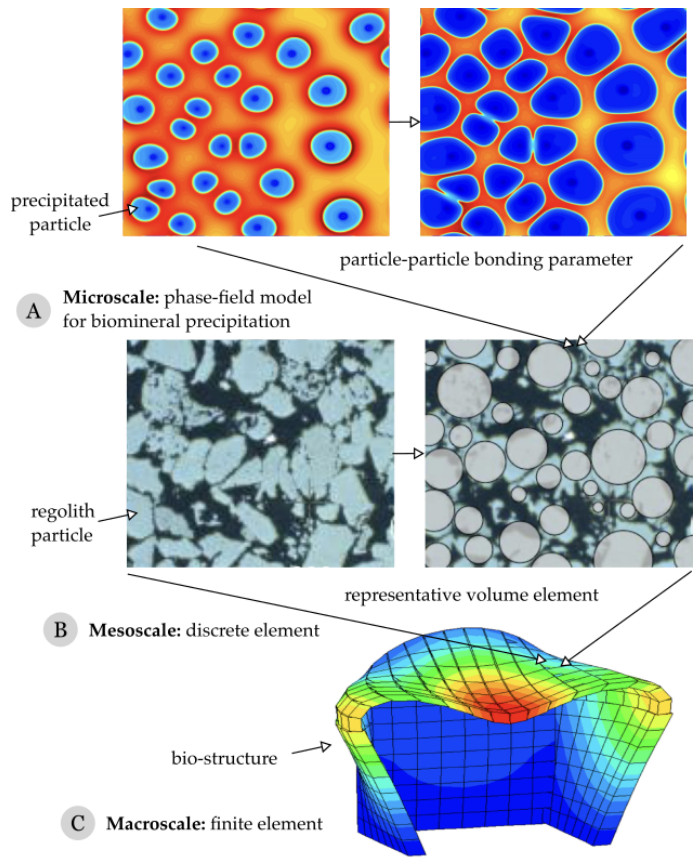

**Figure 11.** A multiscale numerical simulation model is being developed for the project. (a) The evolution of biomineral precipitation among regolith particles is simulated by a phase-field model. (b) The results generated by the microscale model are used as particle-particle bonding parameters and implemented into a mesoscale model. (c) The stochastic upscaling to macroscopic scale can be obtained via homogenization techniques.

As future work, digital twinning will be a perfect tool to achieve real-time monitoring about how Martian construction behaves and evolves. The digital twin will not only provide users with data for monitoring but also use artificial intelligence and machine learning to provide instructions for users and actuators to alter the construction process in real-time and therefore most efficiently produce the structures with desired properties.

To assist the digital twin technology, a physics-based numerical simulation framework is extremely useful. Numerical modeling of the self-growing process often presents significant challenges. Self-growing materials exhibit biomineralization processes at the sub-micron level and create bio-structures with dimensions of meters. Microbes present metabolic reactions in the range of seconds, promote crystal nucleation and growth in the lapse of hours, and sustain their interaction with the surrounding for years. Hence, a multiscale numerical simulation model is needed to address this issue, as shown in Fig. 11. The evolution of biomineral precipitation among regolith particles based on microbial growth and environmental condition is simulated by a phase-field model [21]. The results generated by the microscale model are used as particle-particle bonding parameters and implemented into a three-dimensional mesoscale model built on discrete element methodology [22-24]. The stochastic upscaling to macroscopic scale can be obtained via homogenization techniques [25].

## ACKNOWLEDGEMENTS

This study was funded by the NASA Innovative Advanced Concepts (NIAC) program of the National Aeronautics and Space Administration (NASA) under the grant number of 80NSSC23K0584. This study was also partially funded by the Young Faculty Award (YFA) program of the Defense Advanced Research Projects Agency (DARPA) under the grant number of D22AP00154. The views, opinions, and/or findings expressed are those of the authors and should not be interpreted as representing the official views or policies of the Department of Defense or the U.S. Government. Erin C. Carr was supported by Postdoctoral Research Fellowships in Biology (PRFB) program of National Science Foundation (NSF) under the award number of 2209217.

## CONFLICT OF INTEREST

There are no conflicts of interest.

## DATA AVAILABILITY STATEMENT

The datasets generated and supporting the findings of this article are obtainable from the corresponding author upon reasonable request.

## REFERENCES

1. Scott, A., Oze, C., Hughes, M.W. Magnesium-based cements for Martian construction. J. Aero. Eng. 2020; 33: 4020019.
2. Wan, L., Wendner, R., Cusatis, G. A novel material for in situ construction on Mars: experiments and numerical simulations. Construction and Building Materials 2016; 120: 222-231.
3. Alexiadis, A., Alberini, F., Meyer, M.E. Geopolymers from lunar and Martian soil simulants. Adv. Space Res. 2017; 59; 490-495.
4. Ednie-Brown, P. bioMASON and the speculative engagements of biotechincal architecture. Archit. Des. 2013; 83: 84-91.
5. Dosier, G.K. Composition, tools, and methods for the manufacture of construction materials using enzymes. US8951786B1.
6. Bayer, E., McIntyre, G. Method for producing rapidly renewable chitinous material using fungal fruiting bodies and product made thereby. US20090307969A1.
7. Dikshit, R., Gupta, N., Dey, A., Viswanathan, K., Kumar, A. Microbial induced calcite precipitation can consolidate martian and lunar regolith simulants. PLoS One 2022; 17: e0266415.
8. The National Aeronautics and Space Administration. Could Future Homes on the Moon and Mars Be Made of Fungi? https://www.nasa.gov/directorates/spacetech/niac/2018_Phase_I_Phase_II/Myco-architecture_off_planet
9. Zhang, X., Fan, X., Li, M., Samia, A., Yu, X. Study on the behaviors of fungi-concrete surface interactions and theoretical assessment of its potentials for durable concrete with fungal-mediated self-healing. Journal of Cleaner Production 2021; 292: 125870.
10. Luo, J., Chen, X., Crump, J., Davies, D.G., Zhou, G., Zhang, N., Jin, C. Interactions of fungi with concrete: significant importance for bio-based self-healing concrete. Construction & Building Materials 2018; 164: 275-285.
11. Li, T., Li, C.T., Butler, K., Hays, S.G., Guarnieri, M.T., Oyler, G.A., Betenbaugh, M.J. Mimicking lichens: incorporation of yeast strains together with sucrose-secreting cyanobacteria improves survival, growth, ROS removal, and lipid production in a stable mutualistic co-culture production platform. Biotechnol. Biofuels 2017; 10: 55.
12. Jiang, L., Li, T., Jenkins, J., Hu, Y., Brueck, C.L., Pei, H., Betenbaugh, M.J. Evidence for a mutualistic relationship between the cyanobacteria Nostoc and fungi Aspergilli in different environments. Appl. Microbiol. Biotechnol. 2020; 104: 6413.
13. Cannon, K.M., Britt, D.T., Smith, T.M., Fritsche, R.F., Batcheldor, D. Mars global simulant MGS-1: a Rocknest-based open standard for basaltic Martian regolith simulants. Icarus 2019; 317: 470-478.
14. Eichler, A., Hadland, N., Pickett, D., Masaitis, D., Handy, D., Perez, A., Batcheldor, D., Wheeler, B., Palmer A. Challenging the agricultural viability of Martian regolith simulants. Icarus 2021; 354: 114022.
15. Menon, R.R., Luo, J., Chen, X., Zhou, H., Liu, Z., Zhou, G., Zhang, N., Jin, C. Screening of fungi for potential application of self-healing concrete. Scientific Reports 2019; 9: 2075.
16. Padan, E., Bibi, E., Ito, M., Krulwich, T.A. Alkaline pH homeostasis in bacteria: new insights. Biochim. Biophys. Acta 2005; 1717: 67.
17. Blanco-Rivero, A., Leganes, F., Fernandez-Valiente, E., Calle, P., Fernandez-Pinas, F. mrpA, a gene with roles in resistance to Na+ and adaptation to alkaline pH in the cyanobacterium Anabaena sp. PCC7120. Microbiology 2005; 151: 1671.

18. Hoffmann, D., Gutekunst, K., Klissenbauer, M., Schulz-Friedrich, R., Appel. J. Mutagenesis of hydrogenase accessory genes of Synechocystis sp. PCC 6803. Additional homologues of hypA and hypB are not active in hydrogenase maturation. FEBS J. 2006; 273: 4516.

19. Ushimaru, T., Nishiyama, Y., Hayashi, H., Murata, N. No coordinated transcriptional regulation of the sod-kat antioxidative system in Synechocystis sp. PCC 6803. J. Plant Physiol. 2002; 159: 805.

20. Rokaya, N., Carr, E.C., Wilson, R.A., Jin, C. Lichen-mediated self-growing construction materials for habitat outfitting on Mars. Materials Today. Under Review.

21. Gránásy, L., Rátkai, L., Tóth, G.I., Gilbert, P.U.P.A., Zlotnikov, I., Pusztai, T. Phase-field modeling of biomineralization in mollusks and corals: microstructure vs formation mechanism. JACS Au 2021; 1: 1014-1033.

22. Cusatis, G., Pelessone, D., Mencarelli, A. Lattice Discrete Particle Model (LDPM) for failure behavior of concrete. I: Theory. Cement and Concrete Composites 2011; 33: 881-890.

23. Cusatis, G., Mencarelli, A., Pelessone, D., Baylot, J. Lattice Discrete Particle Model (LDPM) for failure behavior of concrete. II: Calibration and validation. Cement and Concrete Composites 2011; 33, 891-905.

24. Jin, C., Buratti, N., Stacchini, M., Savoia, M., Cusatis, G. Lattice Discrete Particle Modeling of fiber reinforced concrete: experiments and simulations. European Journal of Mechanics A/Solids 2016; 57: 85-107.

25. Li, W., Rezakhani, R., Jin, C., Zhou, X., Cusatis, G. A multiscale framework for the simulation of the anisotropic mechanical behavior of shale. International Journal for Numerical and Analytical Methods in Geomechanics 2017; 41: 1494-1522.